# Air hydrodynamics of the ultrafast laser-triggered spark gap

E. W. Rosenthal[1,*], I. Larkin[1], A. Goffin[1], T. Produit[2], M. C. Schroeder[2], J.-P. Wolf[2], and H. M. Milchberg[1]

(1) Institute for Research in Electronics and Applied Physics, University of Maryland, College Park, MD 20742, USA
(2) Dept. of Applied Physics, University of Geneva, Chemin de Pinchat 22, 1211 Geneva 4, Switzerland

**ABSTRACT**

We present space and time resolved measurements of the air hydrodynamics induced by ultrafast laser pulse excitation of the air gap between two electrodes at high potential difference. We explore both plasma-based and plasma-free gap excitation. The former uses the plasma left in the wake of femtosecond filamentation, while the latter exploits air heating by multiple-pulse resonant excitation of quantum molecular wavepackets. We find that the cumulative electrode-driven air density depression channel initiated by the laser plays the dominant role in the gap evolution leading to breakdown.

## I. INTRODUCTION

Considerable work has been done over the past several decades investigating the triggering of high voltage (HV) gas discharges by intense laser pulses. Spark gap discharges are used in widespread applications including HV surge protection and power switching, high energy laser triggering, and as ignition sources in combustion engines. The theory of spark-gap discharges is rich in basic physics and has been discussed at length in the literature[1-10]. Spark gaps rely on acceleration of free electrons between the cathode and anode by the gap electric field, driving further ionization by collisional avalanche ionization. In the conventional picture, breakdown starts with the development of one or more 'streamers', i.e. avalanche-ionization induced protrusions of charge, which under the action of additional resistive heating of the gas and consequent lowering of neutral gas density, create a higher conductance channel bridging the cathode and anode. Laser heating of the intra-gap gas can enable control of the discharge current path[11]. The use of low energy ultrafast laser pulses can improve this control by generating, via multiphoton or field ionization, a continuous extended length of low density plasma[12]. Extended focal volumes can be generated by optical elements such as cylindrical lenses or axicons[12] or by relying on nonlinear self-guiding by femtosecond filamentation[13,14]. In the case of filamentation in air, on-axis electron densities are typically

---

[*] *currently at Naval Research Laboratory, Washington D.C., 20375, USA*

≲ $10^{16}$ cm$^{-3}$,[15] constituting only ~0.1% fractional ionization at atmospheric density. Few-nanosecond Q-switched lasers, by contrast, can generate higher plasma densities through electron avalanche, but longitudinally extended and contiguous energy deposition is a challenge. The use of double pulse schemes[12,16] or picosecond lasers[17] have been proposed as solutions providing higher density contiguous plasmas.

Regardless of the pulsewidth used, laser triggering of HV discharges in past work has depended on gas ionization by the laser, with the discharge initiated either by the newly conductive channel enabled by the plasma[18], or by the reduced gas density channel driven hydrodynamically by the gas heating[19,20,21], where in the latter case the lower density reduces the breakdown threshold electric field[22]. For femtosecond pulses, because of the relatively low plasma densities and conductivity generated, it has been proposed that hydrodynamic response and on-axis density reduction is the primary mechanism responsible for discharge initiation[20,21]. Among the things we demonstrate in this paper is that a density depression generated with little to no ionization is equally effective in initiating a discharge.

Early work by Loeb[6-8] and Meek[9] explained HV breakdown discharges in terms of streamer formation. The role of streamers and the associated phenomena of leaders and corona discharge have generated much discussion in the femtosecond laser discharge literature on both long and short spark gaps in air. Several groups have determined that corona generation and leader formation are important for filament-guided HV breakdowns in air[20,23-25]. Gordon *et al.* demonstrated a mode of discharge which can proceed without the aid of streamers, where electric field-driven resistive heating by free electrons in the gap increases the temperature and plasma lifetime, leading to breakdown[26]. Schmitt-Sody *et al.* also demonstrated streamer-free HV discharges triggered by > 2ps duration pulses[27].

In all experiments using laser pre-ionization of spark gaps, it is clear that the electric field-driven evolution of the inter-electrode plasma and gas *before* breakdown is crucial in determining the characteristics of the breakdown itself. However, to our knowledge, the relative roles of the conducting plasma and the hydrodynamic gas density reduction in promoting breakdown have not been assessed. In this paper, we perform space- and time-resolved measurements of the plasma and gas evolution in a high voltage electrode gap at times after the application of an ultrashort laser pulse or pulse train up to the point of breakdown. We find that the cumulative electrode-driven air density depression channel initiated by the laser pulse plays the dominant role in the gas evolution leading to breakdown.

## II.  EXPERIMENTAL SETUP

The spark gap consists of two hemispherical tungsten electrodes of radius $a_{elec}$ = 1.27 cm spaced 3 mm–10 mm apart, with 2 mm diameter axial holes for entrance and exit of the heating laser pulse or pulse train, plus an interferometric probe pulse (see Fig. 1(a)). To generate pulse trains, single pulses from a Ti:Sapphire laser (λ=800nm) were first passed through a nested interferometer[28] ('pulse stacker') which generates eight replica pulses, with the inter-pulse delays controlled by motorized translation stages (~10 fs step size). For single pulse experiments, all but one of the pulse stacker arms was blocked. The pulse or pulse train was then passed through an adjustable grating compressor allowing control of the pulsewidth. Inserting the pulse stacker upstream of the compressor avoided nonlinear distortion in the stacker's beamsplitting optics. The laser was then axially focused through the electrode holes (using a $f$ = 50 cm lens at $f/45$), giving a confocal parameter of $2z_0$ = 4 mm and $1/e^2$ intensity radius $w_0$ = 23 µm, with the beam waist placed midway between the electrodes. In general, the axial extent of the gas excitation was longer than $2z_0$ owing to the onset of self-focusing and filamentation, as discussed later.

As shown in Fig. 1(a), the electrodes were connected in parallel with a C=4.4 nF capacitor bank, which was charged through a 1 kΩ resistor up to +30 kV by a DC HV power supply (Spellman High-Voltage model SL30PN10). The diode, inductor, and capacitor near the power supply act as an RF choke to shunt to ground any strong transients from the spark gap breakdown. A current measurement circuit (inside the green dashed box) is inserted in series with the spark gap ground electrode for some of our measurements. Figure 1(b) shows numerical solutions of Laplace's equation giving the on-axis electric field $E_z(z)$ between the electrodes for a range of electrode spacings and a nominal gap voltage of 10 kV. The two short vertical black lines on each curve indicate the front faces of the two hemispherical electrodes, with the electric field quickly decreasing inside the electrode central holes. For electrode gaps $\gtrsim$ 4 mm, $E_z(z)$ is reasonably uniform between the electrodes.

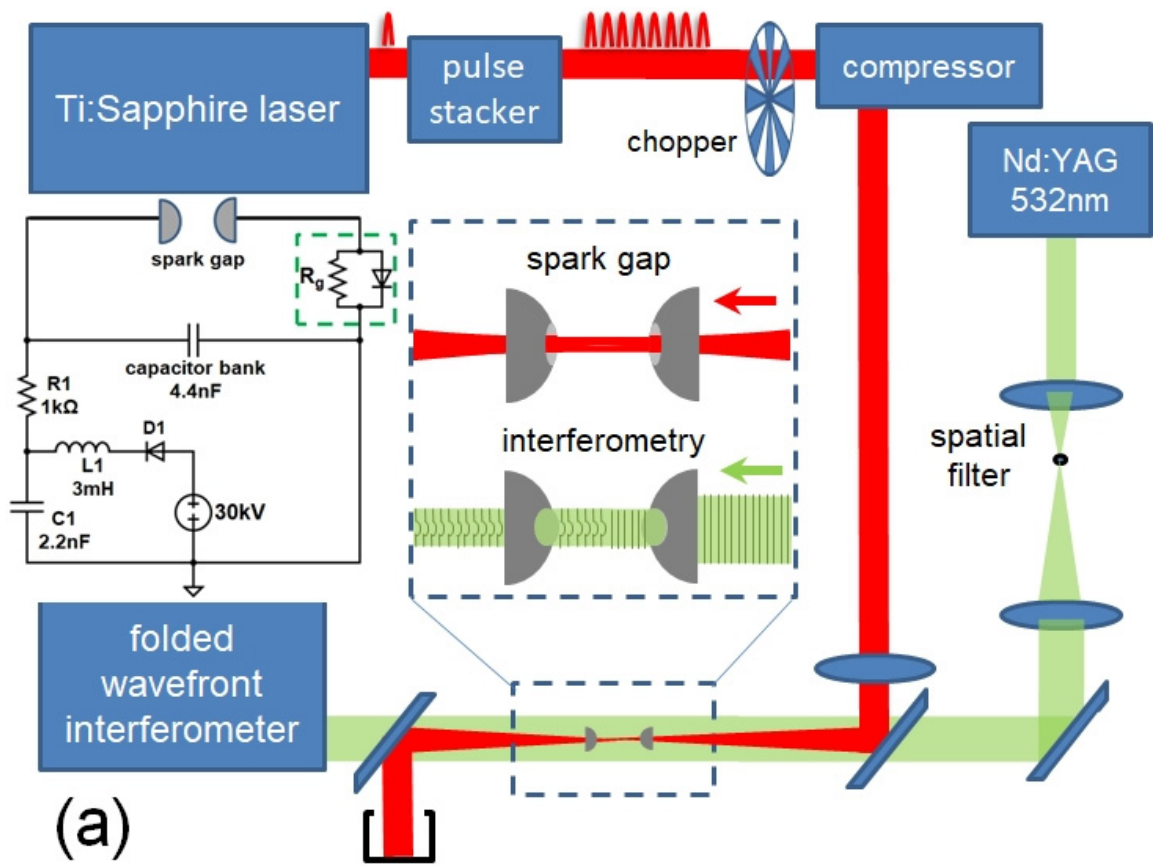

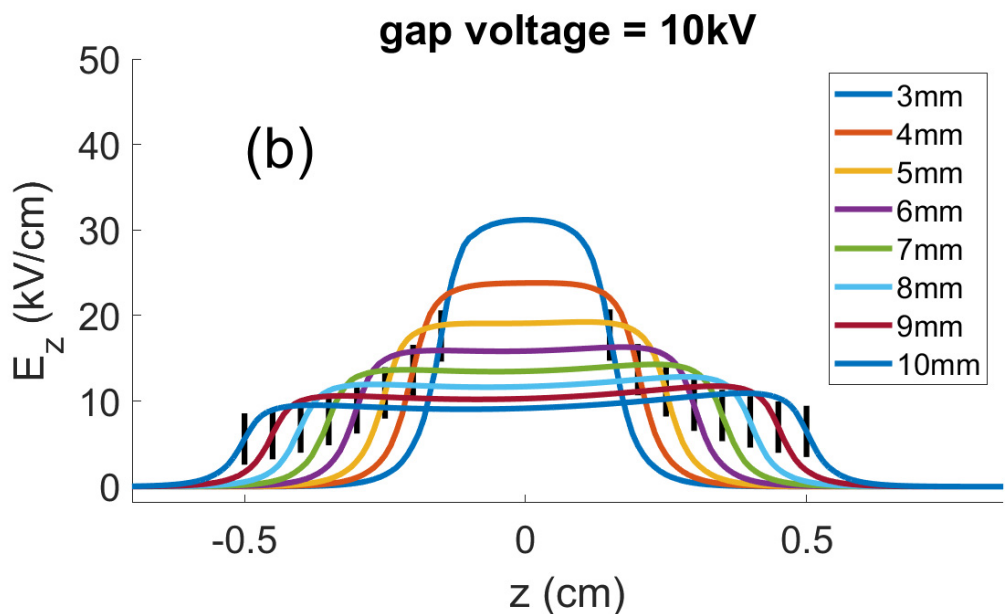

**Figure 1. (a)** Optical setup for investigating gas density dynamics initiated by a λ=800nm femtosecond laser pulse or pulses in a high voltage spark gap. The magnified view shows the femtosecond excitation pulse and λ=532 nm, 10 ns probe pulse propagating axially through the spark gap electrodes. Arrows depict the direction of propagation. Also shown is the spark generation circuit, with the green box depicting an auxiliary current monitor. **(b)** Simulated on-axis axial field $E_z$ for a range of electrode separations for 10 kV gap voltage. Fields for different gap voltages are obtained by linear scaling. The two vertical black bars on each curve indicate the front faces of the hemispherical electrodes.

The gas and plasma evolution between the electrodes was monitored by a variably delayed interferometric probe pulse (λ=532 nm, 10 ns) electronically synchronized (~1 ns jitter) and co-propagating with the femtosecond air excitation pulse(s) (see Fig. 1(a)). The probe pulsewidth and timing jitter were small compared to the onset timescale of breakdown (see below). The interaction region was end-imaged through the hole in the positive electrode onto a folded wavefront interferometer, with the object plane adjustable. Interferometric background images (femtosecond pulse off) were taken on every shot by passing the pulse(s) through an optical chopper before the compressor. The probe beam was cleaned by a spatial filter prior to the interaction region, producing smooth, low noise phase fronts. With use of the chopper, our single shot interferometric measurements were limited to a noise floor of < 40 mrad. Extraction of interferometric phase $\Delta\varphi(\mathbf{r}_\perp)$ was performed as in ref. [29], yielding refractive index perturbation profiles $\Delta n(\mathbf{r}_\perp)$ axially averaged over the gap width, where $\mathbf{r}_\perp$ is a transverse coordinate with respect to the spark gap axis.

## III. EXPERIMENTAL RESULTS

Figure 2 shows a time sequence of air refractive index perturbation profiles $\Delta n(\mathbf{r}_\perp)$ following application of a 65 µJ, 100 fs FWHM laser pulse to a 5.5 mm electrode gap for (a) 0V and (b) 17 kV/cm applied to the gap. Based on measurements and simulations in our prior work[29-31], the profiles in (a) are explained as follows. When a 50-100 fs laser pulse is focused into air, energy is deposited primarily through optical field ionization and non-resonant rotational Raman excitation of the air molecules (the laser bandwidth is not wide enough for vibrational Raman excitation)[32]. The laser-produced plasma recombines to the neutral gas on a <10 ns timescale[15], while the excited molecular rotational wavepacket collisionally decoheres on a ~100 ps timescale[32]. Owing to the finite thermal conductivity of the surrounding neutral gas, the deposited laser energy remains contained in a narrow radial zone, but is repartitioned into the translational and rotational degrees of freedom of the neutral gas[29]. For femtosecond filamentation in atmosphere, the radial zone has approximately the filament core radius of $a_{core} \sim 50$ µm[15,29]. The result is an extended region of high pressure at temperatures up to a few hundred K above ambient[30]. The onset of this pressure spike is much faster than the acoustic timescale of the gas, $a_{core}/c_s \sim 100$ ns, where $c_s \sim 3.4 \times$ $10^4$ cm/s is the air sound speed, so that it impulsively launches a cylindrical single-cycle acoustic wave ~100 ns after the filament is formed[31], as seen in Fig. 2(a). Later panels show that by ~1 μs, the acoustic wave has long since left the filament region, leaving a density depression ('density hole') at elevated temperature and in pressure equilibrium with the surrounding gas. Over ~100 μs to ~1 ms, the density hole decays by thermal diffusion.

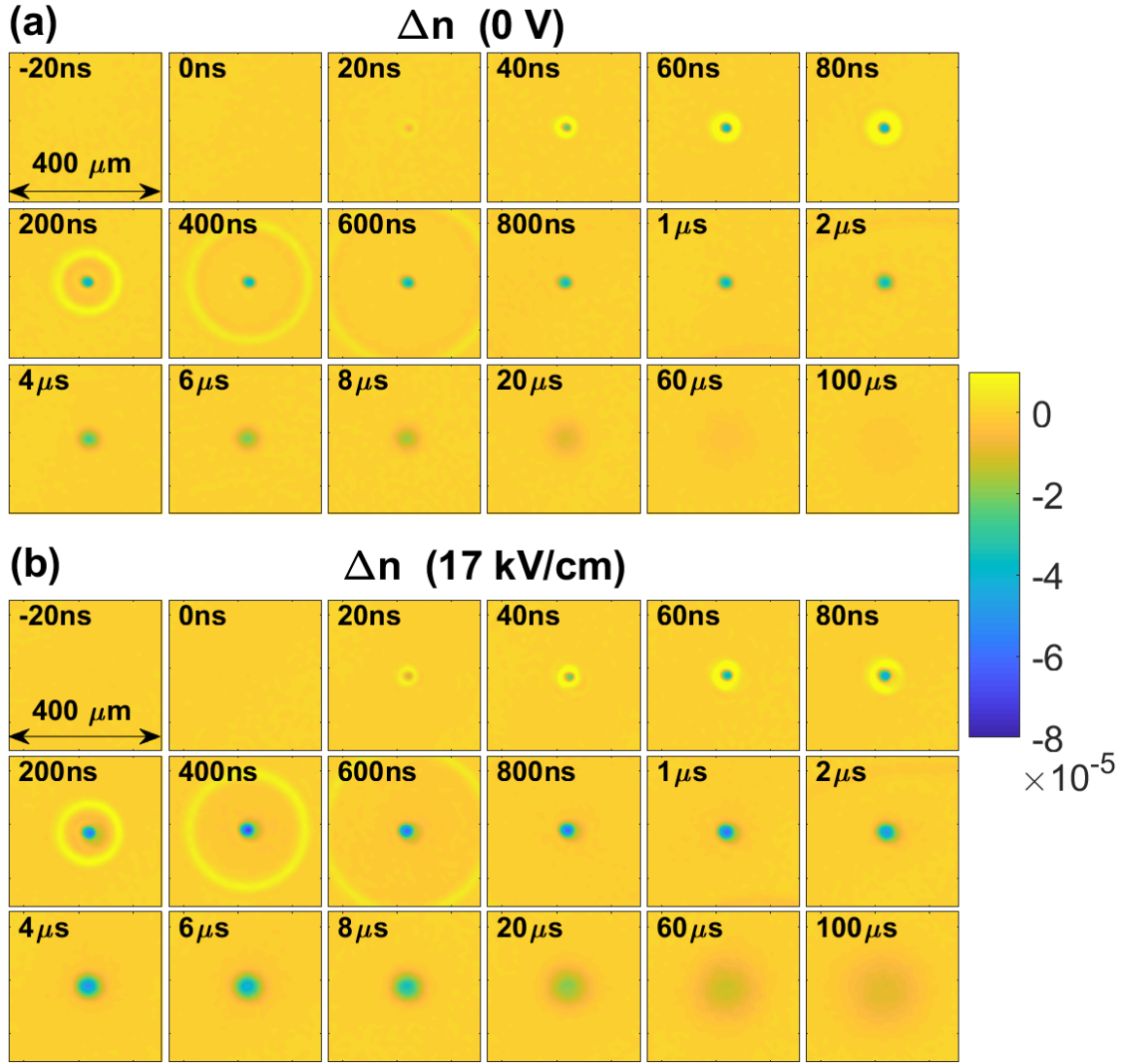

**Figure 2. (a)** Evolution of refractive index shift profiles $\Delta n(\mathbf{r}_\perp)$ at a sequence of probe delays following heating at $t = 0$ by a single 100 fs, 65 µJ laser pulse in the spark gap with zero gap field. The outward propagating yellow ring, seen in the frames at delays < 600ns, is a single-cycle acoustic wave. **(b)** Same measurement as (a), but with 17 kV/cm gap field. In the case of HV applied across the gap, the on-axis density hole is observed to deepen and widen relative to the 0 V case at all delays.

After pressure equilibrium has been achieved, the 'area' of the density hole profile is a proxy for the laser energy deposited per unit length in the gas, as shown in ref. [33],

$$\varepsilon_{abs} = -\frac{c_v T_0 \rho_0}{k(n-1)} \int \Delta\varphi(\mathbf{r}_\perp) \, d^2\mathbf{r}_\perp \qquad (1)$$

where $c_v$ is the specific heat of air at constant volume, $\rho_0$ and $T_0$ are the ambient gas density and temperature, $k = 2\pi/\lambda$ is the probe wavenumber, and $n$ is the gas refractive index. While Eq. (1) was applied in ref. [33] to femtosecond laser-generated density holes, it will also apply to calculating energy deposited by any heating mechanism that is fast compared to thermal diffusion into the surrounding gas, which has a ~millisecond timescale. We use this broader applicability of Eq. (1) in much of the analysis of this paper.

### A. ROLE OF FILAMENT PLASMA AND DENSITY HOLE IN HIGH VOLTAGE BREAKDOWN

We first assess the roles of the laser produced plasma and the gas density depression in the high voltage breakdown process. A first set of experiments was performed in which air density holes of the same depth were generated, either with or without initial plasma. In the case of a single filamenting pulse that generates plasma in the usual manner, the pulse energy was chosen (22 μJ) to produce an on-axis density hole depth $\Delta N/N_0 \sim 3\%$ at 1 μs delay after the pulse, where $N_0$ is the background air density and $\Delta N$ is the on-axis density reduction. In the plasma-free case, we achieved the same $\Delta N/N_0 \sim 3\%$ hole depth at 1 μs delay by using an 8-pulse sequence of 12.5 μJ pulses (below the ionization threshold of the oxygen molecule) from the pulse stacker to rotationally heat the air's nitrogen molecules. The inter-pulse timing in the sequence was adjusted to ~ 8.3 ps (the rotational revival time of $N_2$) in order to maximize the rotational wavepacket excitation and air heating[34]. On the basis of oxygen's ionization rate intensity dependence, $\propto I^8$, we expect single pulse excitation to produce at least $(22~\mu J/12.5~\mu J)^8 \sim 90 \times$ more plasma than the pulse sequence.

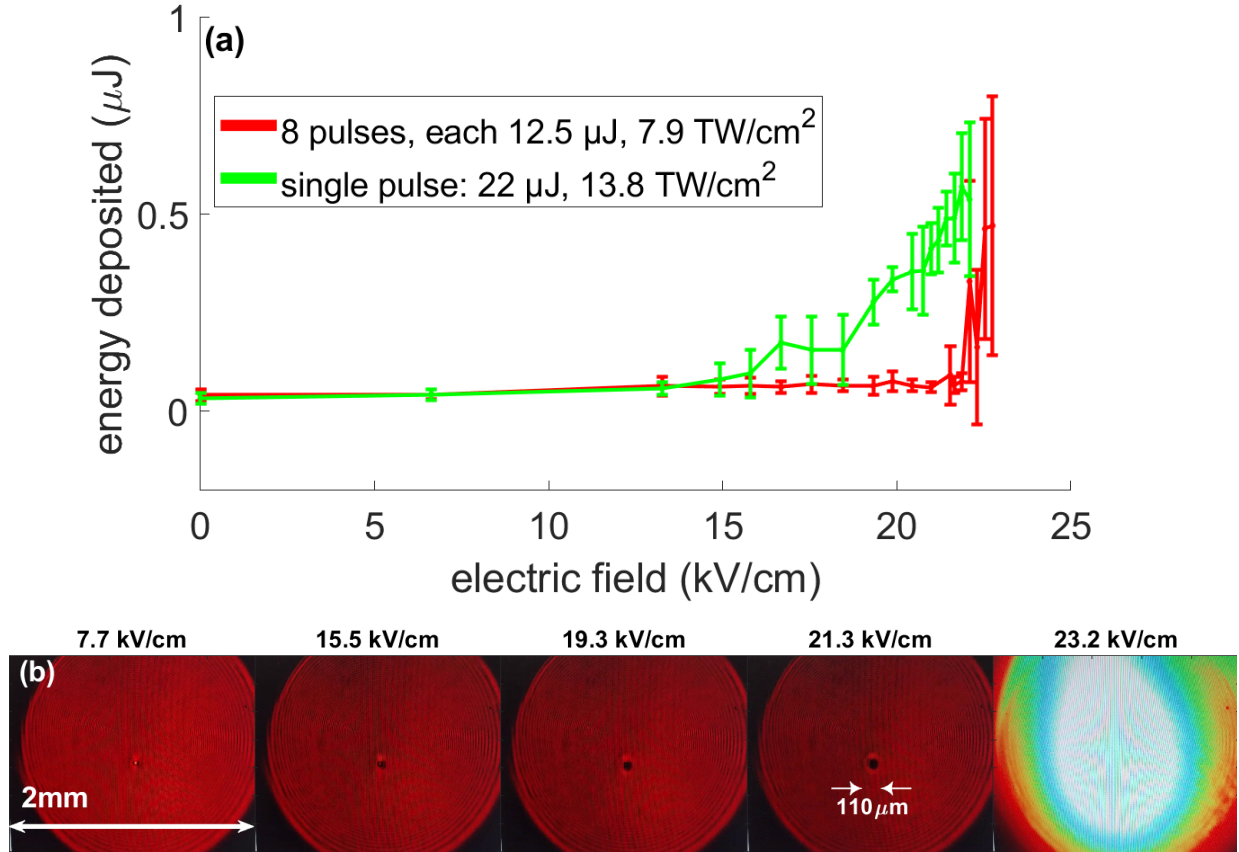

**Figure 3. (a)** Energy deposited in the intra-gap air as a function of gap field for the case of an initial plasma (green curve, single 22 µJ laser pulse) and the case of little to no initial plasma (red curve, 8 pulse sequence with 12.5µJ/pulse). The electrode spacing is 4 mm. For the green curve, a single laser pulse formed a plasma filament between the electrodes. For the red curve, the air in the electrode gap was heated via $N_2$ rotational excitation by the resonant 8-pulse sequence. In both cases, the initial on-axis density hole depth was $\Delta N/N_0 \sim 3\%$ at a delay of 1µs. **(b)** End-view shadowgrams (through positive electrode) of the gas density hole initiated by single 22 µJ pulse and subsequently deepened/widened by increasing gap fields. The edge in the red images is the electrode central hole edge. Probe delay = 3µs. The 23.2 kV/cm frame shows an end view of the breakdown flash. The breakdown is centered on the density hole.

Figure 3(a) shows, for the plasma and near plasma-free cases, the energy deposited in the gas between the electrodes (calculated using Eq. (1) from the gas density hole measured at 1 µs probe delay) as a function of applied gap field. The gap field used here and in all other figures is the average electric field $\langle E_z \rangle$ between the electrodes calculated from Fig. 1(b). Each point is an average over 25 consecutive laser shots, while the error bars correspond to the standard deviation. The plotted points terminate where breakdown occurs. Below ~13 kV/cm, the energy absorbed in the gas (~0.05 μJ) in both cases is consistent with the $\Delta N/N_0 \sim 3\%$ density hole imprinted by the single pulse or pulse train. Above ~13 kV/cm, there is increasing gas heating in the single pulse (plasma) case, consistent with electron impact ionization and resistive heating driven by the high voltage. At the ~13 kV/cm threshold, $eE_{gap}\lambda_{mfp} \sim 1$ eV for a mean free path in air $\lambda_{mfp} \sim 0.5$ μm[35,36]. This is sufficient for electrons to reach several eV over multiple collisions or in the tail of the distribution, enough energy to surmount the nitrogen vibrational $^2\Pi_g$ shape resonance peaking past ~2 eV [1]. We speculate that because of this vibrational energy sink, gap fields below ~13 kV/cm are unable to accelerate electrons sufficiently to accumulate the ~12-15 eV needed for impact ionization of $O_2$ and $N_2$. Both the additional carriers and the higher energy electrons then heat the air via elastic and inelastic collisions, rapidly deepening the density hole and increasing $\lambda_{mfp}$ until the onset of breakdown at ~22 kV/cm, as seen in the green curve. In both plasma-free and plasma cases, however, the breakdown threshold is ~22-23 kV/cm, and occurs roughly where the energy absorbed by the gas (as measured by the density hole volume) is comparable at ~0.5μJ. This suggests that the density hole is the main factor in setting the breakdown threshold. To the extent that pre-existing free electrons are involved, their acceleration in the gap field serves mainly to heat the air to generate the density hole. In the nominally plasma-free case (red curve), pre-breakdown gas heating appears to have occurred, but mainly at fields just below the breakdown level. In another view of the dynamics induced by a single pulse, Fig. 3(b) shows the deepening and widening of the density hole with increasing gap field, until breakdown occurs at ~23 kV/cm.

In the next set of experiments, we examined the effect of the energy of a single filamenting (plasma-producing) laser pulse on the onset of breakdown. Figure 4 plots energy deposition, determined via Eq. (1), as a function of gap field (gap length 5.5 mm) for four laser pulse energies. It is seen that for higher pulse energy, more energy is deposited in the gap and the gap breakdown field (where the curves terminate) is reduced.

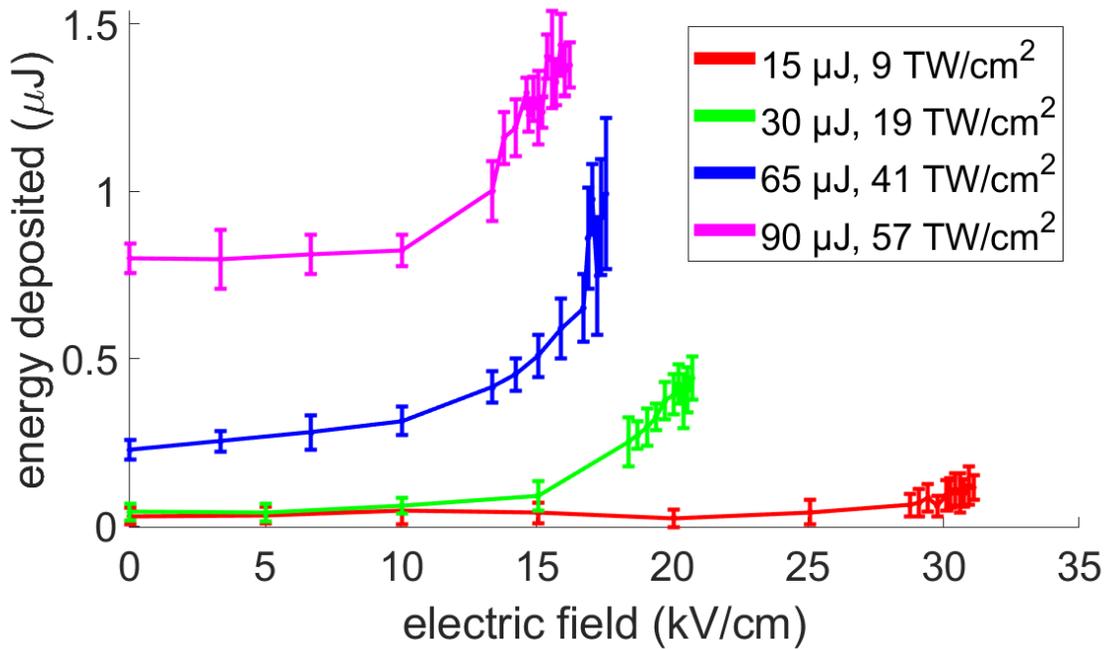

**Figure 4.** Energy deposited in the intra-gap air as a function of the input laser pulse energy and gap field for a 5.5mm gap. The curves terminate at the point of breakdown.

At fields below the sharper upturn of each curve, the energy deposition is seen to slowly increase or stay roughly constant. For each curve, the upturn is associated, as in Fig. 3, with the onset of sufficient electron acceleration for impact ionization of $O_2$ and $N_2$ along with increasing gas heating and widening/deepening of the density hole, which increases channel conductivity. The higher initial electron densities generated by higher laser energies serve to more quickly establish the density hole volume for onset of breakdown, which occurs at lower gap fields.

### B. EFFECT OF SPARK GAP ELECTRODE SEPARATION

Figure 5(a) demonstrates the dependence of air heating in the gap (by a single 75 µJ pulse) as a function of gap electric field for a range of electrode separations (4mm-10mm). Each curve in Fig. 5(a) terminates at the breakdown field near ~15 kV/cm. For the smaller gaps, the effective breakdown field is slightly larger (~16 kV/cm) because, as seen in Fig. 1(b), the field peaks over a smaller fraction of the gap width. For the longer gaps, the peak field occupies a larger fraction of the gap; the breakdown field converges to ~14-15 kV/cm.

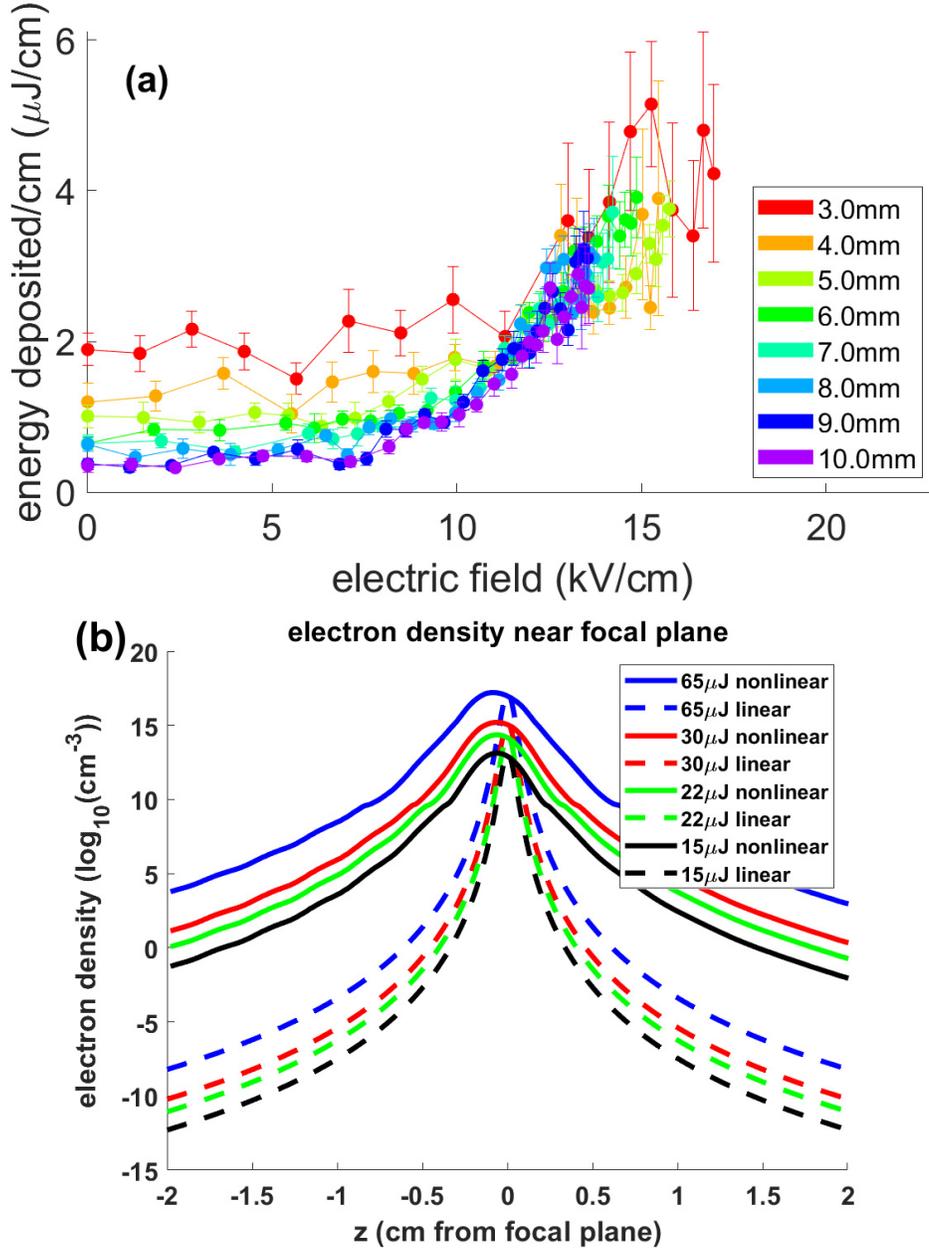

**Figure 5. (a)** Energy deposited per unit length after single 75 µJ laser pulse excitation for a range of gap fields and gap separations. **(b)** Propagation simulation showing the plasma-extending effect of filamentation. The laser propagates left to right. By comparison, the multiphoton ionization yield of a linearly propagating pulse is many orders of magnitude lower away from the linear beam waist. It is seen that filament peak electron density occurs slightly upstream of the linear beam waist consistent with self-focusing and filamentation.

In all of these runs, the laser vacuum confocal parameter is $2z_0 \sim 4$ mm centred on the gap, with spot size $w_0 = 23$ µm. However, filamentary propagation is responsible for extended plasma generation, increasing the electron density away from the vacuum beam waist by many orders of magnitude. This is seen in the filamentation simulations of Fig. 5(b), which are compared to the $\propto (I(z))^8$ ionization yield of a linearly propagating pulse. These simulations were performed with a GPU implementation of the unidirectional pulse propagation equation, which includes the full molecular response of air[37,38]. The plots of Fig. 5(a) show that for low fields, the energy deposition per unit length remains roughly constant, and is higher for smaller gaps. This is consistent with laser energy deposition from the filament plasma, whose axial average electron density is higher for short gaps (see Fig. 5(b)). Only once the field increases to ~8 kV/cm, the curves for all gaps begin to converge as resistive heating (by electrons driven by the gap field) begins to dominate.

### C. TIME DEPENDENCE OF INTRA-GAP GAS HEATING

Without applied HV, after the filament-heated air sheds its single cycle acoustic emission and achieves pressure equilibrium (as seen in Fig. 2(a) for $t > \sim 200$ ns), the density hole decreases in depth and widens as thermal diffusion occurs to the surrounding air. With the application of HV across the gap, the gas is heated continuously after the initial filament energy deposition (as seen by the widening and deepening density hole in Fig. 2(b)). Figure 6 shows plots of the gas heating as a function of time in a 5.5mm gap for a range of voltages for a fixed filamenting laser pulse (65 μJ, 100 fs). The highest voltage was intentionally kept just below the breakdown threshold. By $t = 200$ ns (indicated by the vertical black line), the acoustic wave has just been shed from the density hole. It is only after that time that Eq. (1) can be used to determine absorbed energy. By ~200ns, the gap HV has already heated the air, as indicated by the increasing density hole volume with gap voltage, and heating is shown increasing out to at least ~100 μs. This heating remains predominantly localized to the density hole region imprinted by the initial laser pulse, as seen in Fig. 2(b) and Fig. 3(b).

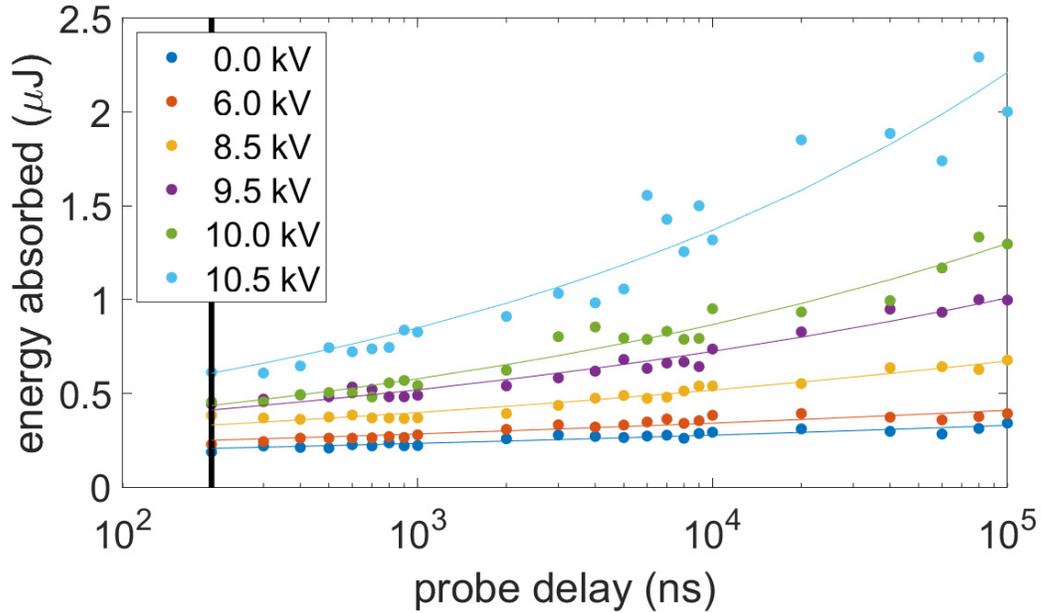

**Figure 6.** Energy absorbed by air vs. probe delay for several gap voltages. The vertical bar corresponds to $t = 200$ ns probe delay, after which Eq. (1) can be applied. The density hole volume is seen to continuously increase out to the maximum probe delay of 100 μs, with an increasingly rapid increase with gap voltage. Each point is an average of 25 shots, and the curves are best fits of the points to $y = at^b$. Even by 200ns, higher gap voltages are seen to do more heating.

### D. PRE-BREAKDOWN GAP CURRENT MEASUREMENTS

To verify that the pre-breakdown air heating is resistively driven by the gap voltage, we inserted an auxiliary current monitor shown by the dashed green box in Fig. 1(a). It consists of a 100MΩ resistor $R_g$ in parallel with a miniature gas discharge tube (LittelFuse CG110) linked to ground from the gap electrode. The discharge tube acts as a shunt to ground for the extremely large transient currents generated at spark gap breakdown. A high voltage probe (Tektronix P6139B) at the top of $R_g$ measures the time-resolved voltage $V_R(t)$ and gap current $I_{gap}(t) = V_R(t)/R_g$ for gap voltage below the breakdown threshold. Results are shown in Fig. 7(a) for a single 22 μJ filamenting laser pulse in a 4 mm gap for gap fields below breakdown. At lower fields, the gap current exhibits a sharp rise at $t = 0$ immediately after application of the laser pulse, followed by a rollover and then an exponential decay, beginning at approximately the vertical dashed black lines. The sharp rise measurement is limited by the probe and oscilloscope frequency response. As the gap field increases, the rollover part of the curves shorten and the response approaches a pure exponential decay following the sharp rise. The $\tau_{1/e} = 115$ μs decay time (see Fig. 8(a) for 22 kV/cm) is in good agreement with an estimate of the characteristic R-C time $R_g C \sim 110$ μs, given by $R_g = 100$ MΩ and the approximate spark gap capacitance $C \sim 4\pi\varepsilon_0 a_{elec}(\frac{1}{2})^{1/3} = 1.1$ pF, where the factor $(\frac{1}{2})^{1/3}$ accounts for the hemispherical electrode. For lower gap fields we see that the actual current transient (to the left of the vertical dashed black lines) is $< \sim 0.25$ ms, much longer than the <10 ns timescale for filament plasma recombination, indicating that a very weak conducting channel is sustained by the gap field until the free carriers in the gap are largely depleted. Figure 7(b) plots the total charge $Q_{gap} = \int_{t_0}^{\infty} I_{gap}(t)dt$ as a function of gap voltage, with the curve terminating at breakdown. Here it clear that the pre-breakdown charge driven across the gap increases with gap field, largely owing to electrode-supplied carriers traversing the increasingly conductive air channel and supplemented by impact ionization. . Measurements of $Q_{gap}$ enable an extreme upper estimate of average electron density $\overline{N_e}$ in the gap during the current pulse: $Q_{gap} \sim \overline{N_e} \pi a^2 L_{gap}$, where $a \sim 50$ μm is the approximate air channel radius (see Fig. 2(b)) and $L_{gap} = 4$ mm is the gap spacing. Using $Q_{gap} \sim 2$ pC gives $\overline{N_e} \sim 4 \times 10^{11}$ cm$^{-3}$. This is orders of magnitude lower than the electron density generated in our femtosecond filaments (see Fig. 5(b)), indicating that electrons from the filament plasma contributes little *directly* to these current pulses.

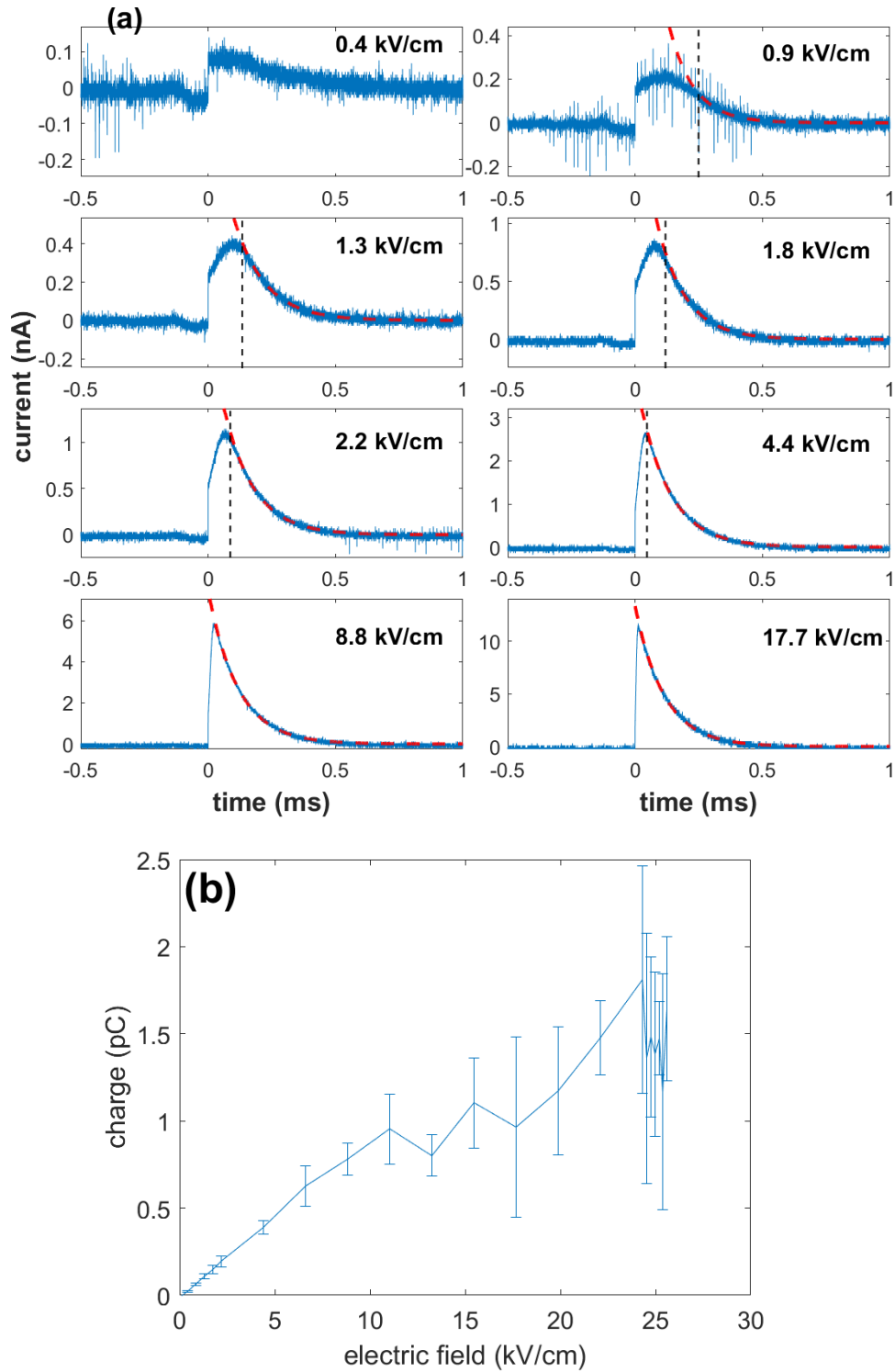

**Figure 7. (a)** Gap current initiated by single 22μJ, 100 fs laser pulse vs. gap field. Exponential decay fits are shown for the > 0.9 kV/cm responses. Impulse response is approached at > ~ 5 kV/cm. Non-impulse time response of < ~0.25 ms occurs to the left of the vertical dashed black lines, where the current transient deviates from the exponential decay. **(b)** Pre-breakdown charge driven across gap vs. gap voltage. The curve terminates at breakdown.

Figure 8 shows results of current measurements made using the two sets of laser parameters of Fig. 3: The single filamenting pulse (22 μJ, 100 fs), and the sequence of eight pulses (12.5 μJ, 100 fs) of lower intensity, separated by the N$_2$ rotational revival interval of 8.3ps. As discussed earlier, the pulse energies were chosen so that the resulting density depression was $\Delta N/N_0 \sim 3\%$ for each case. The impulse response of the gap to the single pulse is shown in Fig. 8(a) for 22 kV/cm gap voltage. Figure 8(b) shows peak current vs. gap field for single (red) and 8-pulse (blue) cases, and a laser-free curve (green) for comparison, where below breakdown, the current is zero. As the gap voltage increases, the peak current for the single pulse case increases faster than the 8-pulse case, owing to its higher initial electron density and increasing air heating and density hole volume (as seen in Fig. 6), increasing the air channel conductance.

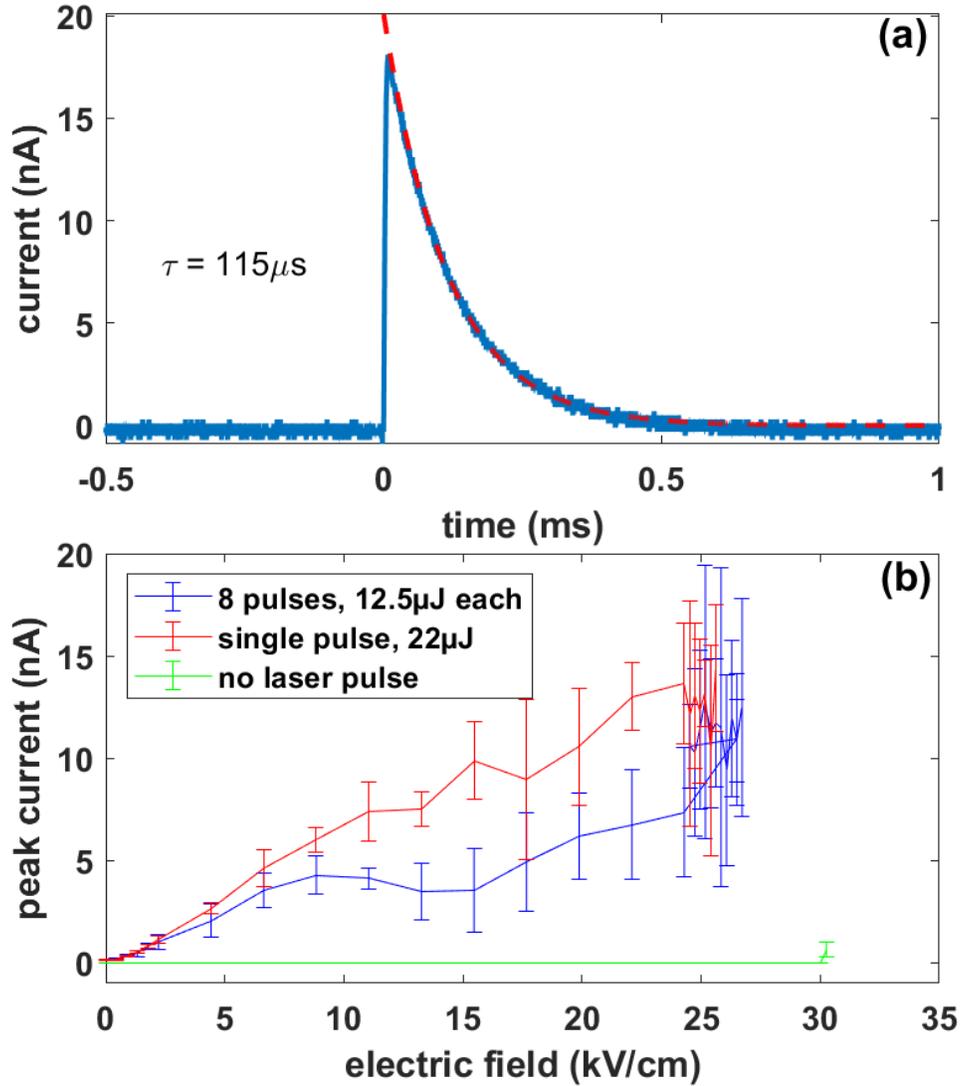

**Figure 8.** Current measurements (below the breakdown threshold) using the green-boxed current monitor depicted in Fig. 1(a). **(a)** Transient gap current at gap field 22 kV/cm induced by single 22 µJ, 100 fs pulse; this is in the impulse response limit (see Fig. 7(a) and discussion). The dashed red line is an exponential fit with $\tau_{1/e} = 115$ µs. **(b)** Peak gap current vs. gap voltage below the breakdown threshold for the single pulse (red, plasma) and 8-pulse (blue, no plasma) cases. The red and blue curves converge and terminate where breakdown occurs at ~25 kV/cm. The no-laser case (green) is shown for comparison; breakdown occurs near ~30 kV/cm.

Even though the 8-pulse train generates less than ~1% of the electron density of the single pulse (see earlier discussion), inspection of Fig. 8(b) shows that its resulting peak current is roughly 50% of the single pulse case until they converge when the gap voltage is close to the breakdown threshold. This strongly suggests that the main role of the laser—whether through generating filament plasma or by rotational excitation—is to initially heat the air and increase the density hole volume to provide enhanced channel conductance; thereafter, free carriers are largely supplied by the gap electrodes. In addition to the air heating, the filament plasma mainly provides initial free carriers in the gap, while for the rotationally heated gas, the initial carriers are likely provided by corona/streamers. In both cases, at sufficiently high gap voltage, the carriers increase in number owing to impact ionization and continue to resistively heat the gas, increasing the density hole volume and conductance. Just before breakdown, the currents are comparable (Fig. 8(b)) as are the density hole volumes (Fig. 3).

## IV. CONCLUSIONS

We measured, for the first time to our knowledge, the spatial and temporal dynamics of gas heating in an ultrashort laser-triggered spark gap prior to breakdown. To elucidate the relative roles of the plasma and the air density depression induced by the laser, we performed longitudinal interferometry through the gap electrodes along with gap current measurements. We find that under all conditions, resistive heating driven by the applied gap field acts to widen and deepen the intra-gap air density channel, leading to eventual breakdown for sufficient field strength. In the case of plasma and density hole generation by a single filamenting pulse, the electrons are driven by the gap field and further heat the air and widen and deepen the hole. In the case of little plasma but comparable density hole generation from an 8-pulse sequence, the lower current initially provided by corona/streamers is preferentially channeled through the higher conductance hole, heating it further. As gap voltage is increased, impact ionization increases the gap current to the point where the current and the density hole volume in both cases are comparable. This leads to similar breakdown thresholds.

To summarize, once a density hole is created between the electrodes, and regardless of how it was formed, it acts as a preferred channel through which subsequent gap field-driven current may flow. This current collisionally heats the channel further, widening and deepening it and increasing conductance and current, leading to eventual breakdown for sufficient gap field. For ease of diagnostic access, we examined relatively short spark gaps; however, we expect this scenario to apply to much longer femtosecond filament-triggered discharges.

## V. ACKNOWLEDGEMENTS

This work is supported by the Army Research Office (W911NF-14-1-0372), Air Force Office of Scientific Research (FA9550-16-1-0121, FA9550-16-1-0284), and the Office of Naval Research (N00014-17-1-2705, N00014-17-1-2778). MCS and JPW gratefully acknowledge financial support from the Schweizerischer Nationalfonds zur Förderung der Wissenschaftlichen Forschung under grant no. 200021-178926. TP and JPW acknowledge the European Union's Horizon 2020 Research and Innovation program under grant agreement no. 737033-LLR.